\documentclass[aps,preprint,groupedaddress,showpacs]{revtex4}
\usepackage[dvips]{graphicx}
\begin{document}

\title{Structural and Electronic Properties of Amorphous and
Polycrystalline In$\rm _2$Se$\rm _3$ Films}

\author{A. Chaiken}\email{chaiken@hpl.hp.com}
\author{K. Nauka}
\author{G.A. Gibson}
\author{Heon Lee}\altaffiliation[Now at ]{Pohang University of
Science and Technology, Pohang, Korea}
\author{C.C. Yang}
\affiliation{Hewlett-Packard\\
  1501 Page Mill Rd.\\Palo Alto, CA 94304}

\author{J. Wu}
\author{J.W. Ager}
\author{K.M. Yu}
\author{W. Walukiewicz}
\affiliation{Materials Sciences Division\\Lawrence Berkeley National
Laboratory\\Berkeley, CA 94720}

\date{\today}

\begin{abstract}
Structural and electronic properties of amorphous and single-phase
polycrystalline films of $\gamma$- and $\kappa$-In$\rm _2$Se$\rm _3$
have been measured.  The stable $\gamma$ phase nucleates homogeneously
in the film bulk and has a high resistivity, while the metastable
$\kappa$ phase nucleates at the film surface and has a moderate
resistivity.  The microstructures of hot-deposited and post-annealed
cold-deposited $\gamma$ films are quite different but the electronic
properties are similar.  The increase in the resistivity of amorphous
In$\rm _2$Se$\rm _3$ films upon annealing is interpreted in terms of
the replacement of In-In bonds with In-Se bonds during
crystallization.  Great care must be taken in the preparation of
In$\rm _2$Se$\rm _3$ films for electrical measurements as the presence
of excess chalcogen or surface oxidation may greatly affect the film
properties.
\end{abstract}

%\keywords{}

\maketitle

\section{INTRODUCTION}

In$\rm _2$Se$\rm _3$ films have been studied as a precursor to
CuInSe$\rm _2$ for solar cells,\cite{noufi} as an optical recording
medium,\cite{hitachi} as a rotary polarizer for
optoelectronics,\cite{ye2} and, in Li-intercalated form, for battery
applications.\cite{julien} The binary phase diagrams for III-VI
compounds are complicated, and the unit cells of the III-VI crystal
structures are large.  This complexity originates from the many
energetically similar ways that trivalent and divalent atoms can
combine to satisfy their bonding requirements.  In thin films a
variety of different In-Se phases may be formed depending on the
deposition method and parameters.  For a given stoichiometry these
phases differ primarily in the way that the cation vacancies are
arranged.

In$\rm _2$Se$\rm _3$ crystallizes in the well-known layered
($\alpha$-phase) or defect wurtzite ($\gamma$-phase)
structures,\cite{ohtsuka} or a recently discovered anisotropic
structure ($\kappa$-phase).\cite{kees} Ye {\it et al.}  discovered
large optical rotary power in what they term the vacancy ordered in
screw form (VOSF) phase.\cite{ye} In layered compounds like
$\alpha$-In$\rm _2$Se$\rm _3$ or InSe, the cation vacancies form a
plane, resulting in weak Se-Se bonding and anisotropic electronic
properties.\cite{jaegermann} In$\rm _2$Se$\rm _3$ and InSe are
anisotropic hexagonal or rhombohedral semiconductors that tend to have
a high resistivity.\cite{yudasaka} In$\rm _4$Se$\rm _3$ on the other
hand is a highly conducting smaller-bandgap orthorhombic
semiconductor.\cite{julien2} The higher conductivity of In$\rm
_4$Se$\rm _3$ is attributed to the presence of In-In bonds, while in
the other In-Se compounds indium bonds only to
selenium.\cite{marsillac2} The interpretation of electronic and
optical data on polycrystalline thin films of the III-VI compounds is
made more difficult by the frequent occurrence of more than one of
these crystalline phases even in perfectly stoichiometric films.

Many different methods have been employed to prepare In$\rm _2$Se$\rm
_3$ films, including coevaporation from elemental sources,\cite{emery}
flash evaporation from the stoichiometric compound,\cite{guesdon}
sputtering,\cite{hitachi}
% electrodeposition,\cite{massaccesi}
% metal-organic chemical vapor deposition\cite{obrien} 
and even annealing of cold-deposited In/Se
multilayers.\cite{sahu,marsillac,oyelaran,lu} In this paper we
describe structural, electronic, and optical characterization of
In$\rm _2$Se$\rm _3$ films deposited by a number of methods and
discuss how small quantities of impurity phases or slight surface
oxidation can have a large impact on transport measurements.  To the
extent that they can be determined, the single-phase properties of
unoxidized polycrystalline In$\rm _2$Se$\rm _3$ films are reported and
compared to values in the literature.

\section{EXPERIMENTAL METHODS}

Indium-selenium films were deposited from the elemental constituents
using either electron-beam sources or Knudsen cells in a chamber with
a base pressure near 1x10$\rm ^{-8}$ torr.  In$\rm _2$Se$\rm
_{3-x}$Te$\rm _x$ ternaries were also studied.  Zn, Si and Ge doping
and Te alloying were accomplished using separate elemental sources.
Substrate materials were either silicon or oxidized silicon, but the
choice of substrates did not affect the films' composition or
morphology.  Deposition rates as measured by a quartz crystal were in
the range of 0.1-1 nm/s and typical thickness was 400 nm.  No
dependence on deposition rate or thickness was noted.  Substrates were
optionally heated before, during or after deposition at temperatures
up to 450~$^{\circ}$C.  The substrate temperature was monitored by a
backside thermocouple in spring-loaded contact.  Substrates that were
not intentionally heated rose to a maximum of 60 $^{\circ}$C during
deposition; the resulting films are called ``cold-deposited.''  For
some samples, thin Al$\rm _2$O$\rm _3$ encapsulation layers were
sputtered from an oxide target after indium selenide deposition.  {\it
Ex situ} post-annealing was performed in a tube furnace with a base
pressure near 1x10$\rm ^{-6}$ torr and plumbed with both 99.998\%-pure
Ar and an Ar-5\%H$\rm _2$ mixture.

Composition of the films was measured using energy-dispersive x-ray
analysis (EDX) and Rutherford backscattering (RBS).  Structural
analysis was carried out using a four-circle x-ray diffractometer
(XRD) and Cu K$\alpha$ radiation.  X-ray diffraction data are shown as
semilog plots to emphasize the presence or absence of secondary
phases.  On selected samples transmission electron microscopy (TEM)
was performed using standard techniques.\cite{jacek}
Cathodoluminescence measurements were performed in a Philips SEM using
both photomultiplier and Ge detectors.  Photoluminescence,
photoreflectance and Raman spectroscopy were performed on some
higher-quality films.  The excitation energy for PL was 2.6 eV.
% The
%time-resolved photoluminescence decay (TRPL) method was used to
%estimate carrier recombination lifetimes and trap densities.
Hall measurements were performed at room temperature using 4-lead and
6-lead geometries in a 1.2 T magnet.  Metallic contacts for electrical
measurements were made from sputtered Mo and were found to be linear
for all but the most resistive films.

\section{RESULTS}

\subsection{Composition and Crystal Structure}

Polycrystalline indium-selenium films were produced using three
distinct methods: coevaporation onto a heated substrate
(``hot-deposited'' films); coevaporation onto an unheated substrate
followed by post-annealing (``cold-deposited'' films); and deposition
of an indium-selenium multilayer onto an unheated substrate, followed
by post-annealing (``multilayer'' films).  The three types of
depositions may be expected to favor different crystallization
sequences and result in different microstructures and defects in the
polycrystalline films.  Post-annealing could be performed either in
the deposition chamber (``in situ'') or in the tube furnace (``ex
situ'').  When Ar-H$\rm _2$ was used as the annealing gas, the two
annealing methods produced identical results, as will be discussed
further below.  Finally, cold-deposited and multilayer films could be
annealed either with or without the presence of an Al$\rm _2$O$\rm _3$
passivation layer.  Below data for uncapped films are presented first.

\begin{figure}
%\begin{center}
%\leavevmode
%\epsffile[37 342 576 758]{phased.ps}
%\end{center}
%\includegraphics*[bb=37 342 576 758]{phased.eps}
\includegraphics*{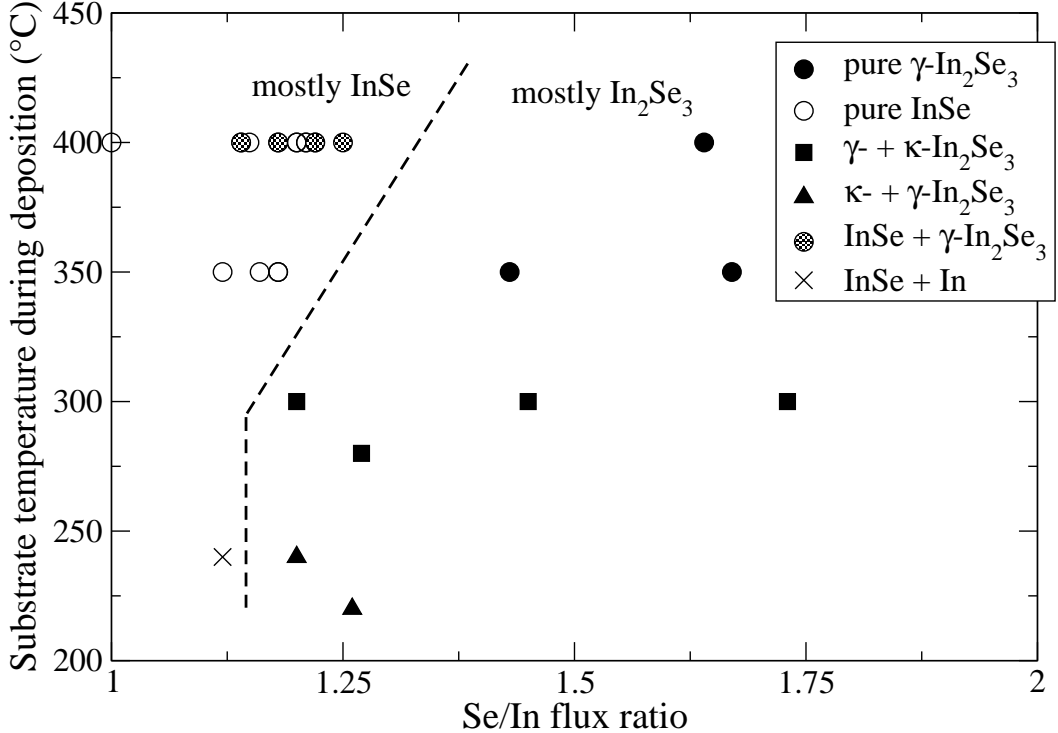}
\caption{\label{phased} Phase diagram of the Indium-Selenium system
showing which compounds are produced as a function of substrate
temperature and Se overpressure.  ``$\gamma$- + $\kappa$-In$\rm
_2$Se$\rm _3$'' means a film that is primarily $\gamma$-In$\rm
_2$Se$\rm _3$ with $\kappa$ as a minority phase.  The dashed line is
the separatrix between the fields where In$\rm _2$Se$\rm _3$ and InSe
predominate.}
\end{figure}

Figure~\ref{phased} shows conditions which produce the various phases
of In$\rm _2$Se$\rm _3$, InSe and In for hot-deposited films.  Phase
purity was determined by x-ray diffraction.  Holding the substrates at
125~$^{\circ}$C to remove water vapor was not sufficient to induce
crystallization, instead producing amorphous films. In general higher
Se fluxes and higher substrate temperatures produced single-phase
$\gamma$-In$\rm _2$Se$\rm _3$, although Se/In flux ratios over 2.0 and
substrate temperatures under 300~$^{\circ}$C resulted in incorporation
of elemental Se into the films.  Lower temperatures resulted in the
formation of the recently discovered $\kappa$ phase, which appears to
be an analog of the $\alpha$ phase with reduced
ordering.\cite{kees,jacek,chris} The $\alpha$ phase was not observed
in films deposited by any of the three methods.

The most surprising feature of Fig.~\ref{phased} is that
stoichiometric In$\rm _2$Se$\rm _3$ was reproducibly formed by growth
under Se-deficient conditions.  Annealing of cold-deposited uncapped
films confirmed what was found in hot-deposited specimens, namely that
large deviations from Se/In = 1.5 resulted in the presence of InSe or
Se secondary phases, but small deviations could be corrected by
annealing.  Both EDX and RBS showed that In or Se excesses as large as
a few percent could be removed by heat treatment either in vacuum or
inert gas.  Ternary films lost both excess Se and Te, with Te/Se
ratios decreasing during annealing, consistent with previous
reports.\cite{elmaliki}

\begin{figure}
%\begin{center}
%sample 57B (hot) and 43A (cold) and 62C (multilayer)
\includegraphics*{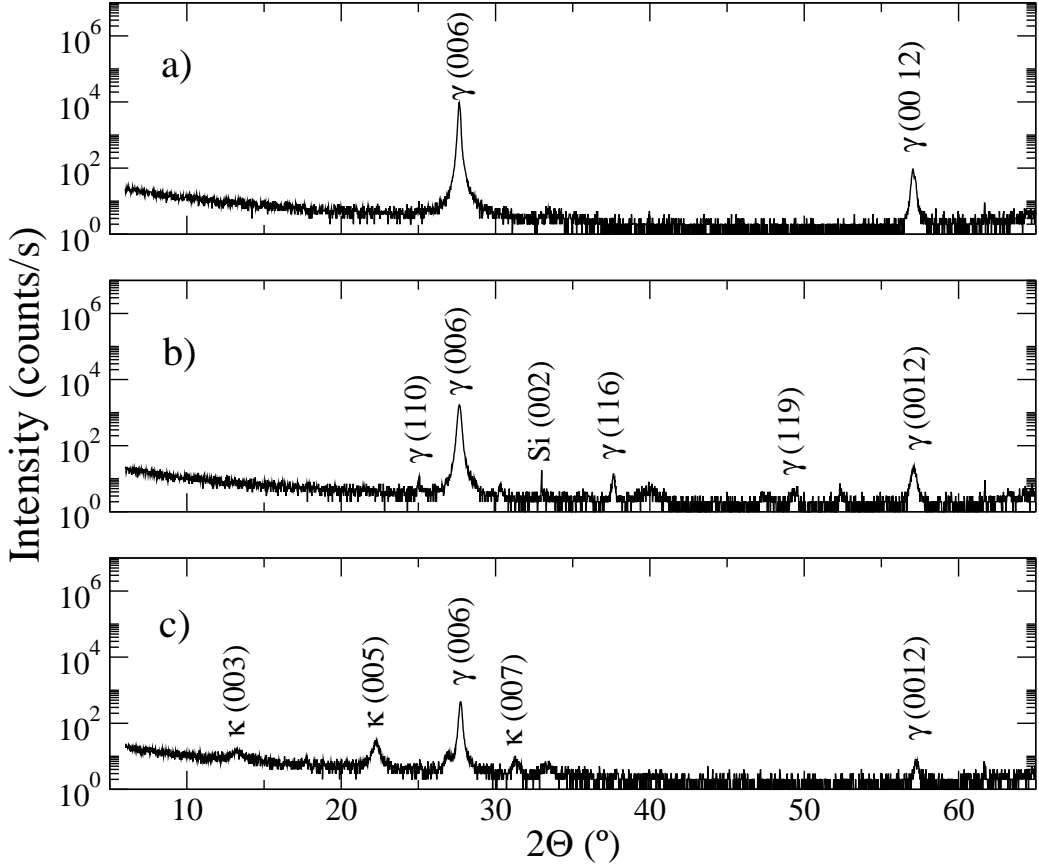}
\caption{\label{gammaxrd} X-ray diffraction spectra for typical In$\rm
_2$Se$\rm _3$ films.  a) A co-evaporated film deposited at
350~$^{\circ}$C.
%with a Se/In flux ratio of 1.67.
b) A co-evaporated,
cold-deposited film 
%made with a Se/In flux ratio of 1.49 and
post-annealed at 350~$^{\circ}$C for 30 minutes in vacuum.  c) An
In$\rm _2$Se$\rm _3$ film synthesized by annealing a
cold-deposited (In 9.2 \AA/Se 15.0 \AA)x145 multilayer {\it in situ}
at 400~$^{\circ}$C for 30 minutes.}
\end{figure}

\begin{figure}
\includegraphics*{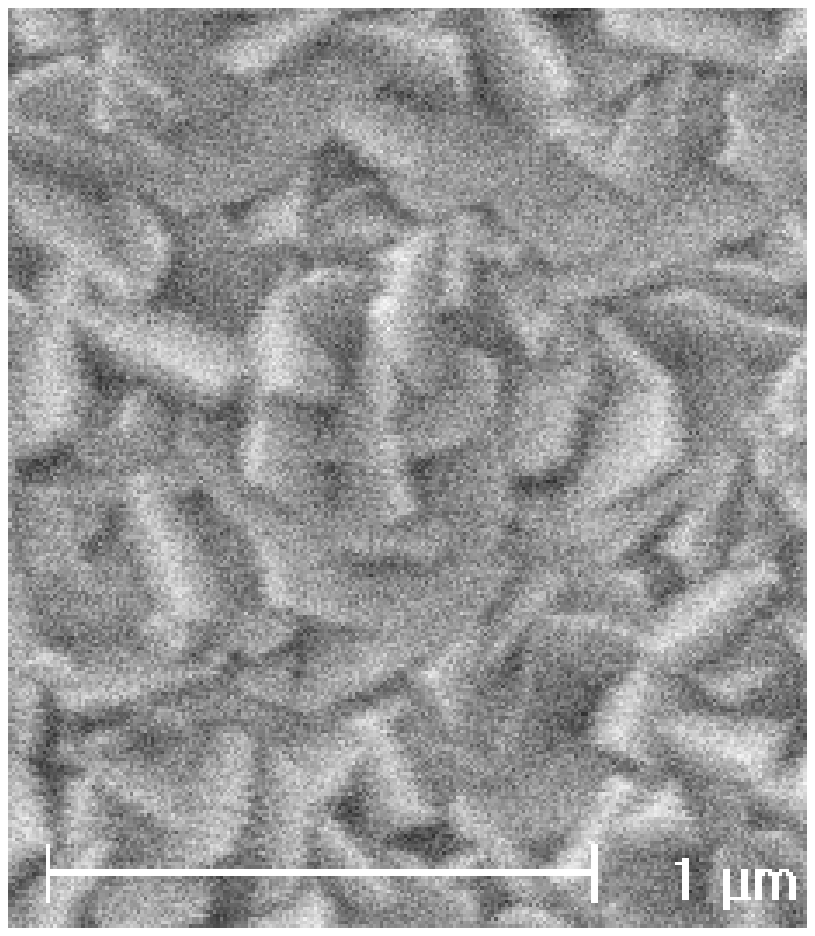}
\includegraphics*{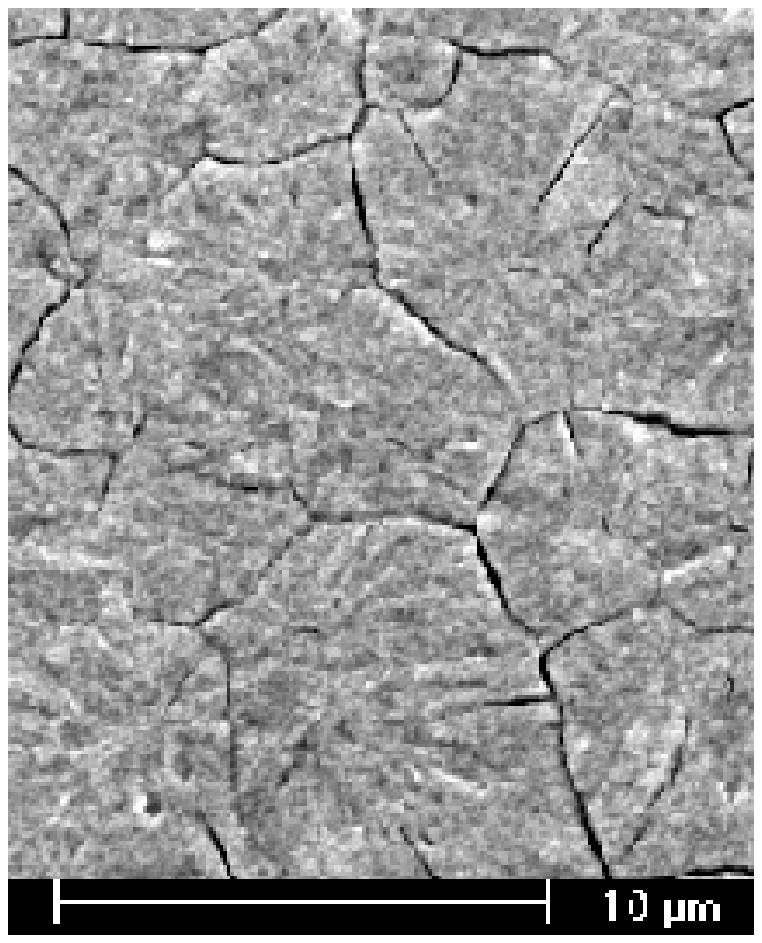}
\caption{\label{gammasem} SEM images of typical In$\rm _2$Se$\rm _3$
films.  a) A co-evaporated film deposited at 350$^{\circ}$C substrate
temperature shows small tilted grains.  b) A cold-deposited uncapped
film {\it ex situ} annealed at 330~$^{\circ}$C for 2 hours in Ar-H$\rm
_2$ shows large, plate-like grains.}
%%%%%%%%%%%%%%%%%%%%%%%%%%%%%%%%%%%%%%%%%%%%%%%%%%%%%%%%%%%%%%%%%%%%%%%%%%%%%%%
%%	\caption{\label{gammasem} SEM images of typical In$\rm _2$Se$\rm _3$ %%
%%	films.  a) A co-evaporated, cold-deposited film made with a Se/In flux%%
%%	ratio of 1.33 and {\it ex situ} annealed at 330~$^{\circ}$C for 2    %%
%%	hours in Ar.  b) A co-evaporated film deposited at		     %%
%%	350$^{\circ}$C with a Se/In flux ratio of 1.43.}		     %%
%%%%%%%%%%%%%%%%%%%%%%%%%%%%%%%%%%%%%%%%%%%%%%%%%%%%%%%%%%%%%%%%%%%%%%%%%%%%%%%
%b) A co-evaporated film deposited at 150$^{\circ}$C with
%a Se/In flux ratio of 1.56 and annealed {\it in situ} at 350
%$^{\circ}$C for 30 minutes.}
\end{figure}

Figure~\ref{gammaxrd}a shows a $\theta$-2$\theta$ XRD spectrum for a
typical hot-deposited pure $\gamma$-In$\rm _2$Se$\rm _3$ film that
contains multiple crystallographic orientations.  In contrast,
Figure~\ref{gammaxrd}b shows the diffraction spectrum of an uncapped
cold-deposited, co-evaporated {\em ex situ} annealed (350~$^{\circ}$C, 30
minutes in vacuum) In$\rm _2$Se$\rm _3$ film.  The single phase,
complete (00$\ell$) orientation and large grains are typical for
cold-deposited, post-annealed films.  X-ray diffraction on {\it in
situ} post-annealed multilayer films (Figure~\ref{gammaxrd}c)
indicates that they are fine-grained and contain some $\kappa$-In$\rm
_2$Se$\rm _3$ crystallites, just like hot-deposited films.
Figure~\ref{gammasem} shows SEM images of the grains in hot- and
cold-deposited co-evaporated films.  Consistent with the x-ray
diffraction results, the hot-deposited film consists of small grains
of mixed orientation, while the post-annealed cold-deposited film is
made up of large, flat (00$l$)-oriented platelets.

%Films prepared by annealing
%In/a-Se multilayers have either large or small grains depending on
%thermal treatment, consistent with previous
%observations.\cite{emziane,lu}

Quite different results have been found for cold-deposited capped
films.  RBS and EDX show that alumina (Al$\rm _2$O$\rm _3$) capping
layers of 7 nm thickness suppress the evolution of volatile species,
with the result that the composition of capped cold-deposited films
did not change during annealing.  Consequently while post-annealed
uncapped films and {\em in situ} annealed films that are subsequently
capped are often single-phase In$\rm _2$Se$\rm _3$, x-ray diffraction
showed that post-annealed capped films typically contained small
amounts of secondary phases, usually InSe or Se.

\begin{figure}
%\begin{center}
%\leavevmode
%sample 36A
%\epsffile[40 315 580 754]{kappasuppress.ps}
%\end{center}
%sample 36A
%\includegraphics*[bb=37 359 576 741]{kappasuppress.ps}
\includegraphics*{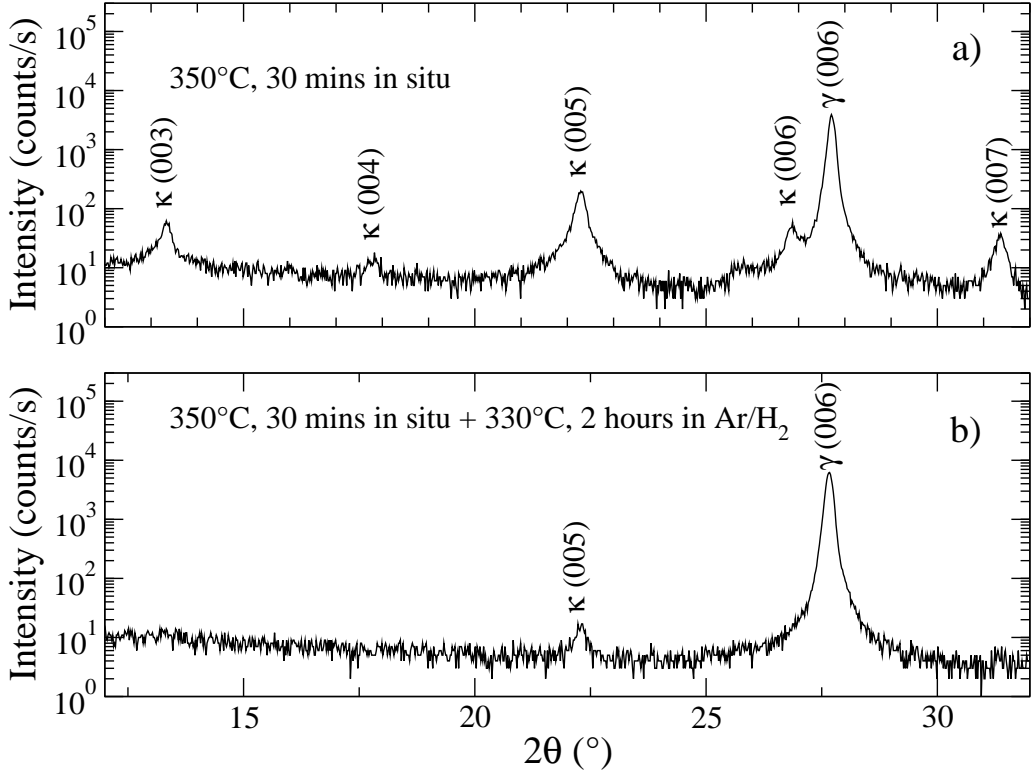}
\caption{\label{kappasuppress}X-ray diffraction spectra of a
co-evaporated, cold-deposited In$\rm _2$Se$\rm _3$ film.  a) After
annealing at 350~$^{\circ}$C for 30 minutes {\it in situ} after the
deposition.  Both $\gamma$ and $\kappa$ phases are present. b) After a
further {\it ex situ} anneal of the same film in Ar/H$\rm _2$ for
330~$^{\circ}$C for 2 hours.  The $\kappa$ phase has been almost
completely suppressed.}
\end{figure}

$\kappa$-In$\rm _2$Se$\rm _3$ occurred in films deposited at substrate
temperatures in the 150 to 250~$^{\circ}$C range, and in films that
were deposited at lower temperatures and then annealed in the vacuum
chamber. {\em Ex situ} annealed cold-deposited films may crystallize
in the $\kappa$ structure when annealed in an Ar-5\%H$\rm _2$
atmosphere, but never when annealed in Ar alone.  These recent
observations lend support to the idea\cite{kees} that surface
oxidation and capping suppress formation of the $\kappa$ phase.  As
shown in Fig.~\ref{kappasuppress}, further annealing of $\kappa$-phase
films slowly transforms them into $\gamma$, suggesting that the
$\kappa$ phase is metastable.  It's not clear if air exposure during
acquisition of the first x-ray data shown in Fig.~\ref{kappasuppress}
played a role in the $\kappa$-$\gamma$ transformation.  Post-annealing
has no effect on the microstructure of hot-deposited $\gamma$-phase
films.

\begin{figure}
%sample 89B
\includegraphics*{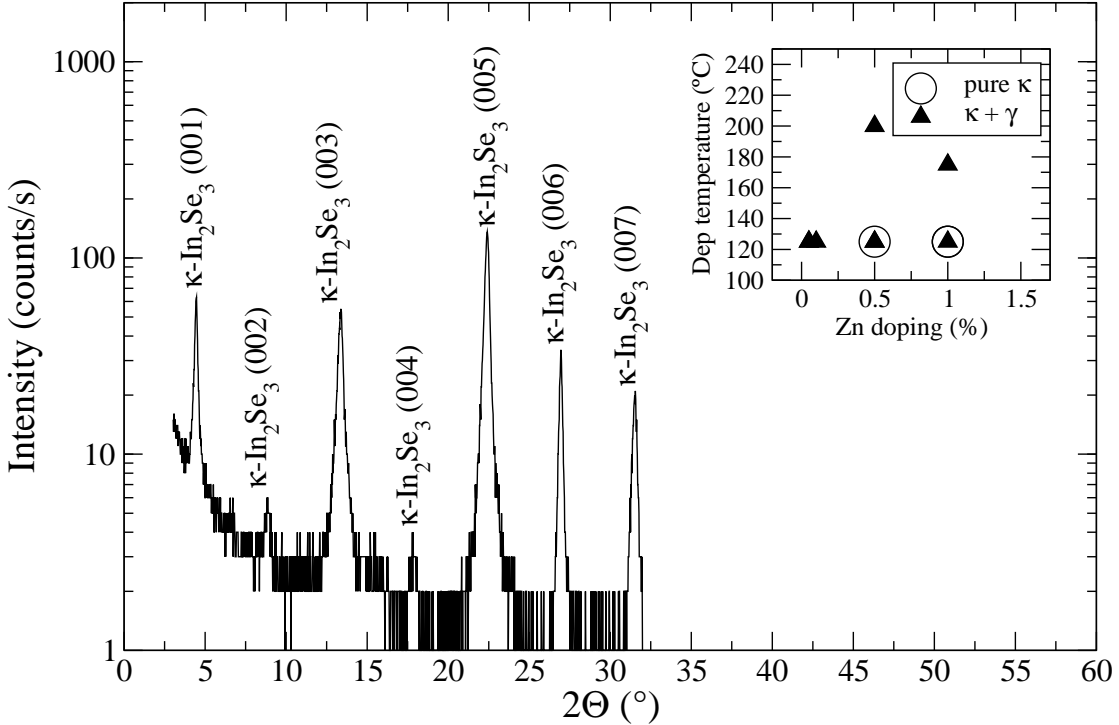}
\caption{\label{kappaxrd}X-ray diffraction spectrum of a film with
nominal composition In$\rm _2$Zn$\rm _{0.01}$Se$\rm _3$ deposited at
125~$^{\circ}$C and then annealed {\it in situ} at 350~$^{\circ}$C for
30 minutes.  The presence of Zn allows formation of pure, well-ordered
$\kappa$-phase.  There were no peaks from the film above 32$^{\circ}$.
The inset shows the deposition temperature-Zn content conditions that
yielded pure $\kappa$-phase films.  All films were annealed to
350$^{\circ}$C {\em in situ} after deposition.}
\end{figure}

Zn doping in the range of 0.5-1\% has permitted stabilization of a
pure $\kappa$-In$\rm _2$Se$\rm _3$ film for the first time.
Reference~\onlinecite{kees} reported that Ag doping stabilized the
$\kappa$ phase but with much broadened XRD peaks.  Some peaks with
similar d-spacings were reported for In$\rm _2$Zn$\rm _{0.4}$Se$\rm
_3$.\cite{haeuseler} The XRD spectrum in Fig.~\ref{kappaxrd} shows
that all (00$l$) peaks appear, in contrast to the $\gamma$-In$\rm
_2$Se$\rm _3$ spectrum, which typically shows only the (006) and (00
12) peaks.  The $\kappa$ spectrum has the unusual feature that odd-$l$
(00$l$) peaks are more intense than even-$l$ ones.  Analysis of TEM
diffraction patterns has yielded lattice constants $a$ = 8.09 $\pm$
0.10 \AA\ and $c$ = 19.85 \AA.\cite{jacek} The latter value is in
excellent agreement with the perpendicular d-spacing found by analysis
of the XRD pattern.

\subsection{Resistivity and Hall Measurements}

\begin{figure}
%\begin{center}
%\leavevmode
%\epsffile[41 338 580 754]{carrvsmob.ps}
%\end{center}
%\includegraphics*[bb=41 338 580 754]{carrvsmob.ps}
\includegraphics*{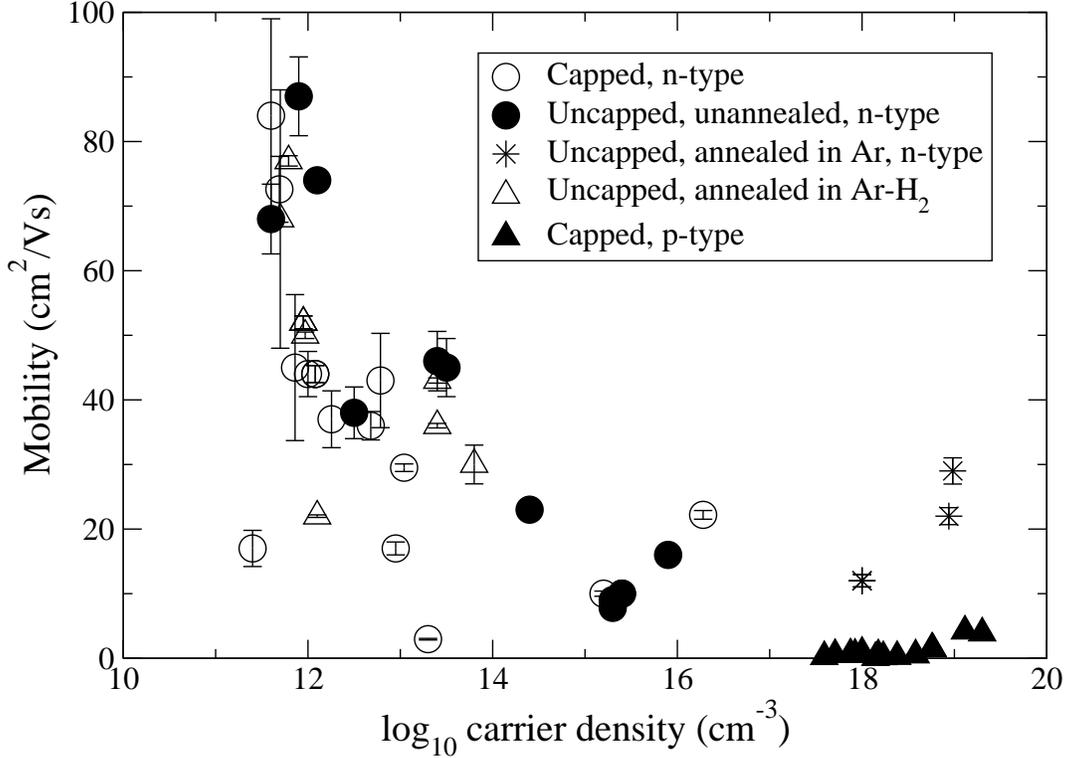}
\caption{\label{carriersvsmobility} A semilog plot of mobility versus
carrier concentration for a number of In$\rm _2$Se$\rm _3$ films shows
that higher carrier densities occur in films with lower mobilities.
The errors originate from field-independent offsets in Hall
measurements.}
\end{figure}

Figure~\ref{carriersvsmobility} shows the results of transport
measurements on a large number of In$\rm _2$Se$\rm _3$ films.  Many
other films were too resistive for successful Hall measurements.
Except for the uncapped n-type films annealed in Ar and the capped
p-type films, all the other data show that lower mobilities are
correlated with higher carrier densities.  X-ray diffraction and RBS
show that all p-type films have excess Se or Te.  Since polycrystaline
Se is itself a relatively conducting p-type semiconductor,\cite{kasap}
it is likely that Se inclusions dominate the transport properties of
these films.  No p-type conduction was ever observed in uncapped films
or in capped films that XRD, EDX and RBS showed to be stoichiometric.
Similarly, the {\em in situ} annealed cold-deposited capped n-type
films with high carrier density and low mobility were found by XRD to
have inclusions of $\kappa$-In$\rm _2$Se$\rm _3$, as discussed further
below.  The presence of InSe as a secondary phase did not affect the
resistivity.

The high conductivity of the uncapped Ar-annealed films in
Figure~\ref{carriersvsmobility} appears to be explained by the
presence of small amounts of In$\rm _2$O$\rm _3$, which is a
degenerate n-type semiconductor.\cite{ali} On a few films very weak
x-ray diffraction peaks matching the published d-spacings of In$\rm
_2$O$\rm _3$ were observed.  An experiment comparing top-contact and
bottom-contact Hall measurements on a highly conducting uncapped film
showed definitively that a surface layer was responsible for its low
resistivity.  The use of Ar:H$\rm _2$ ambient rather than pure Ar
prevented oxide formation.  Unlike the slightly oxidized samples
annealed in Ar, samples annealed in Ar:H$\rm _2$ have characteristics
similar to the uncapped, unannealed films and the {\em in situ}
annealed capped films.

\begin{figure}
%\begin{center}
%\leavevmode
%\epsffile[37 338 576 754]{carrvscomp.ps}
%\end{center}
%\includegraphics*[bb=37 338 576 754]{carrvscomp.ps}
\includegraphics*{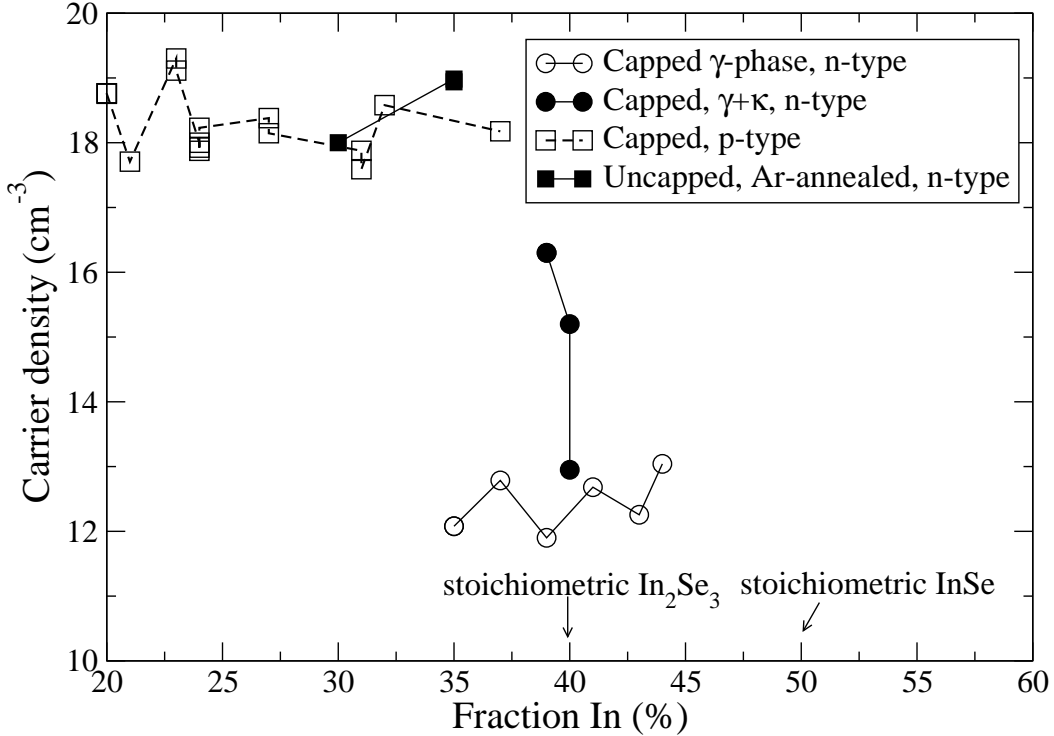}
\caption{\label{carriersvscomposition} The dependence of carrier type
and density on composition illustrates the fact that only capped films
with trapped excess chalcogens are p-type.  Pure $\gamma$-phase films
closer to stoichiometry are n-type with a lower carrier density,
except for uncapped films which have a conducting oxide on the
surface. Films with inclusions of $\kappa$-phase tend to have higher
carrier densities.  All compositions come from RBS or EDX
measurements.  The accuracy is estimated to be $\pm$1\%.}
\end{figure}

The remaining unexplained data points in
Figure~\ref{carriersvsmobility} are in the low carrier density, high
mobility regime.  Apparently these samples exhibit the intrinsic
transport properties of $\gamma$-In$\rm _2$Se$\rm _3$ films, namely
n-type carriers with density $<$ 10$\rm ^{13} cm^{-3}$ and 20 $\le \mu
\le$ 60 cm$\rm ^2$/Vs.  No matter how films were deposited and no
matter what their exact composition, texture or grain size, the
transport properties of crystalline $\gamma$-phase films varied only
slightly, as the data for capped $\gamma$-phase samples in
Figure~\ref{carriersvscomposition} illustrate.

\begin{figure}
\includegraphics*{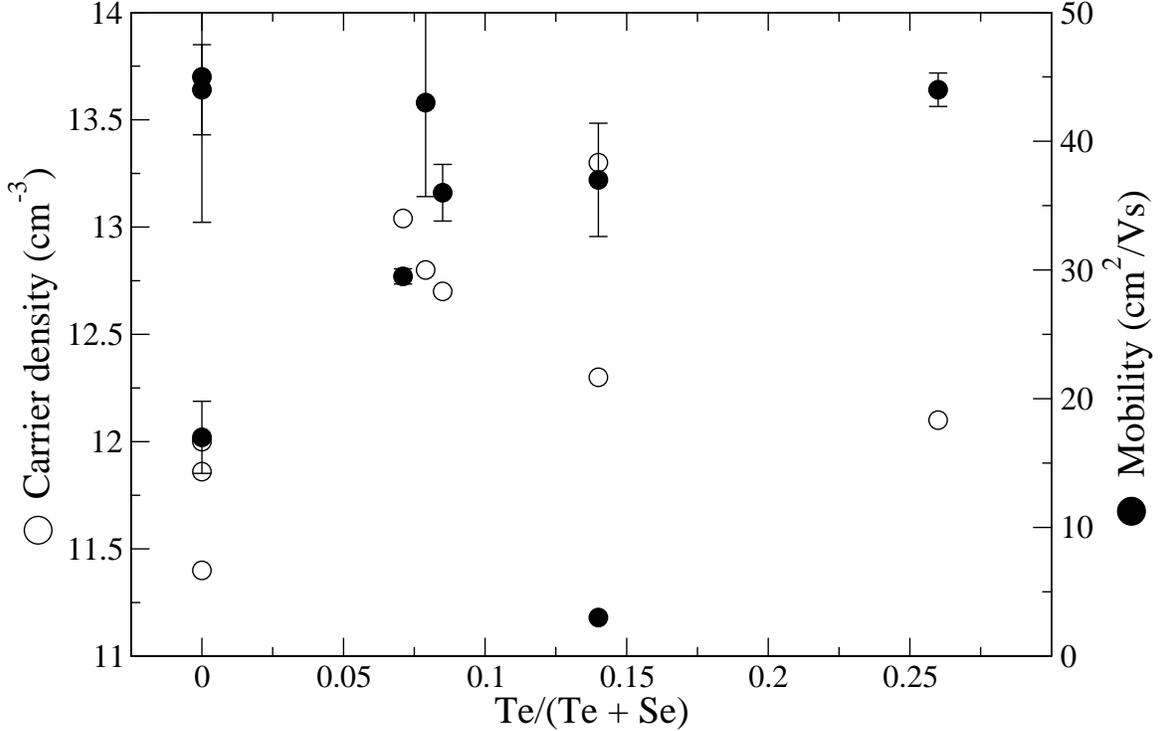}
\caption{\label{vste}Capped n-type In$\rm _2$Se$\rm _{3-x}$Te$\rm _x$
film electrical properties do not show a trend with Te content.
Zero-Te $\kappa$-phase films have been excluded.}
\end{figure}

Attempts at Si and Ge doping and annealing in a Se atmosphere did
little to vary the carrier concentration as long as single-phase
$\gamma$-In$\rm _2$Se$\rm _3$ was preserved.  Tellurium alloying
experiments followed the same trends, namely that capped films were
p-type due to excessive chalcogen, while unoxidized uncapped films had
a low carrier density and higher mobility.  The dependence on Te
content in n-type films with intrinsic behavior is quite weak, as
illustrated by figure~\ref{vste}.  Any variation in properties as a
function of Te doping is smaller than the scatter in the properties of
undoped films.  Overall the properties of In$\rm _2$Se$\rm
_{3-x}$Te$\rm _x$ ternaries were minimally different from those of
In$\rm _2$Se$\rm _3$ binaries with the same total chalcogen/In ratio.

\begin{figure}
\includegraphics*{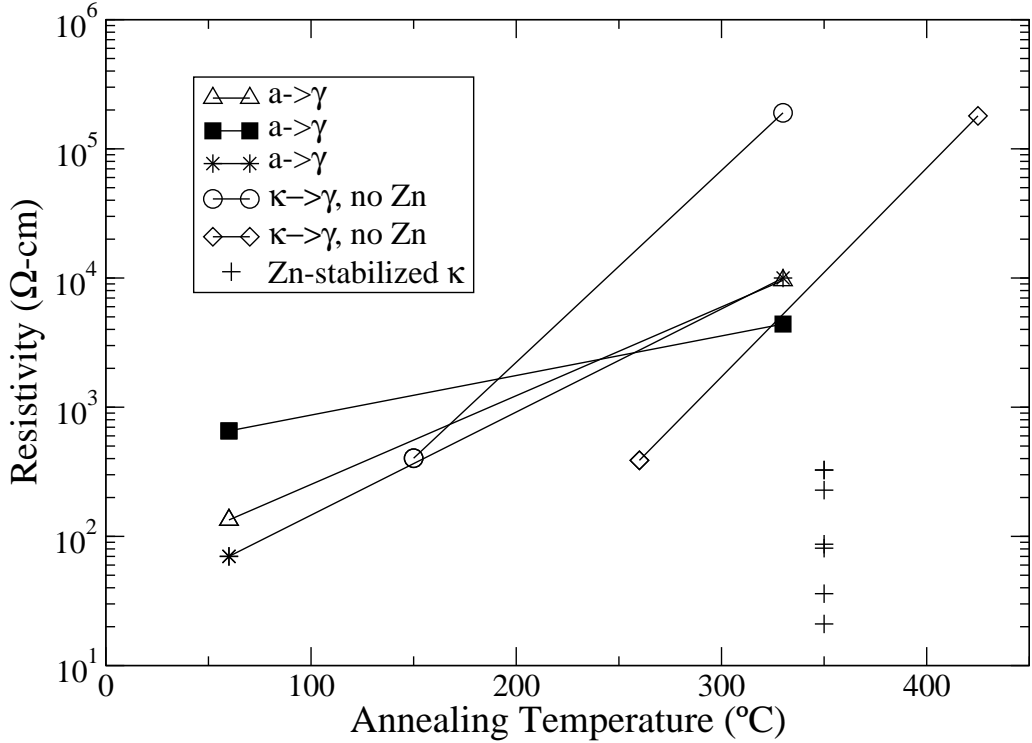}
\caption{\label{anneal} Resistivity versus heat-treatment temperature
for uncapped In$\rm _2$Se$\rm _3$ films.  For pairs of point connected
by a line, the left-hand point is as-deposited, and the right-hand
point is for the same specimen after further {\em ex situ} annealing
in Ar-H$\rm _2$.  The Zn-stabilized samples were deposited at 125
$^{\circ}$C and annealed {\em in situ} to 350 $^{\circ}$C.  Films that
x-ray diffraction showed to be amorphous (indicated as ``a'') actually
had a lower resistivity than well-annealed $\gamma$-phase films
irrespective of the grain size or texture of the polycrystalline
material.  Only $\kappa$-phase films retain a resistivity as low as
the as-deposited state.}
\end{figure}

Films containing $\kappa$-In$\rm _2$Se$\rm _3$ were significantly more
conducting than those containing $\gamma$ phase only.  Well-ordered
single-phase $\kappa$ films made with 0.5-1\% Zn doping have a
mobility 30 $\le \mu \le$ 70 cm$\rm ^2$/Vs and a carrier density of
5x10$\rm ^{14} \le $ n $\le$ 1x10$\rm ^{16}$ cm$\rm ^{-3}$.  Since Zn
substituted on the In site should give p-type carriers, the Zn
impurities are not acting as dopants, but may simply be serving to
stabilize the more conducting $\kappa$ phase.
%47C and 50C
$\kappa$-containing films are not broken out in
fig.~\ref{carriersvsmobility}, but are shown separately in
figs.~\ref{carriersvscomposition} and \ref{anneal}.
Figure~\ref{anneal} demonstrates that conversion of $\kappa$-phase or
amorphous material into $\gamma$ as illustrated in
fig.~\ref{kappasuppress} can dramatically increase the resistivity.
Prolonged annealing may also promote migration and clustering of
defects.  Surprisingly, annealing of well-crystallized nominally pure
$\gamma$-phase uncapped films in an inert atmosphere invariably
resulted in an increase in resistivity even though the x-ray
diffraction spectra did not change.

\begin{figure}
%\begin{center}
%\leavevmode
%\epsffile[43 330 585 770]{activeenergy.ps}
%\end{center}
%\includegraphics*[bb=40 308 579 745]{activeenergy.eps}
\includegraphics*{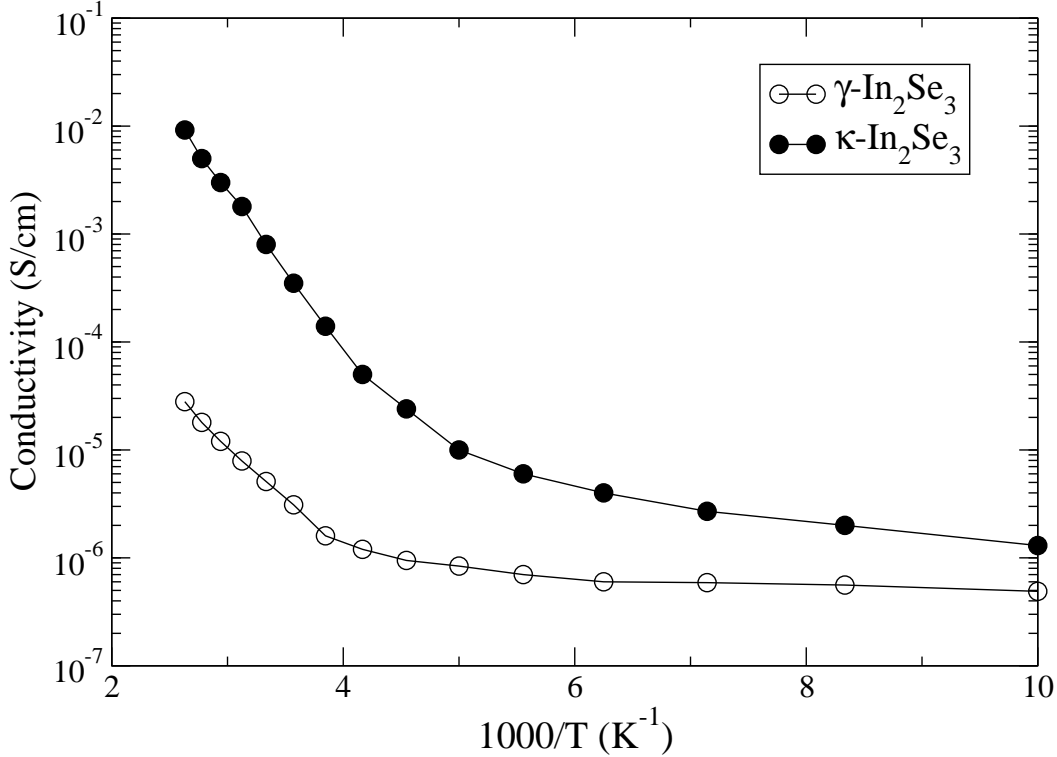}
\caption{\label{activeenergy} The temperature dependence of the
conductivity for typical $\gamma$- and $\kappa$-In$\rm _2$Se$\rm _3$
films shows two different regimes of behavior.}
\end{figure}

Figure~\ref{activeenergy} shows the temperature-dependent conductivity
of typical $\gamma$ and $\kappa$ films.  There are two distinct
temperature regimes.  At temperatures above about 200 K, Arrhenius
behavior is observed with an activation energy of 110 meV for $\gamma$
and 70 meV for $\kappa$.  These activation energies are much less than
the reported bandgaps of 1.8-1.9 eV for the $\gamma$
phase\cite{julien3,kees} and 1.75 eV for $\kappa$,\cite{kees} implying
that the carriers are derived from shallow defects.  The conductivity
saturates at lower temperatures where the carriers are frozen out.
Yudasaka {\it et al.} reported activation energies between 130 and 780
meV depending on heat treatment of the $\gamma$-In$\rm _2$Se$\rm _3$
films.\cite{yudasaka} Bern\`{e}de {et al.} report activation energies
between 0.20 and 0.75 eV depending on temperature of measurement and
heat treatment of the film.\cite{bernede}

\subsection{Spectroscopic Measurements}

\begin{figure}
\includegraphics*{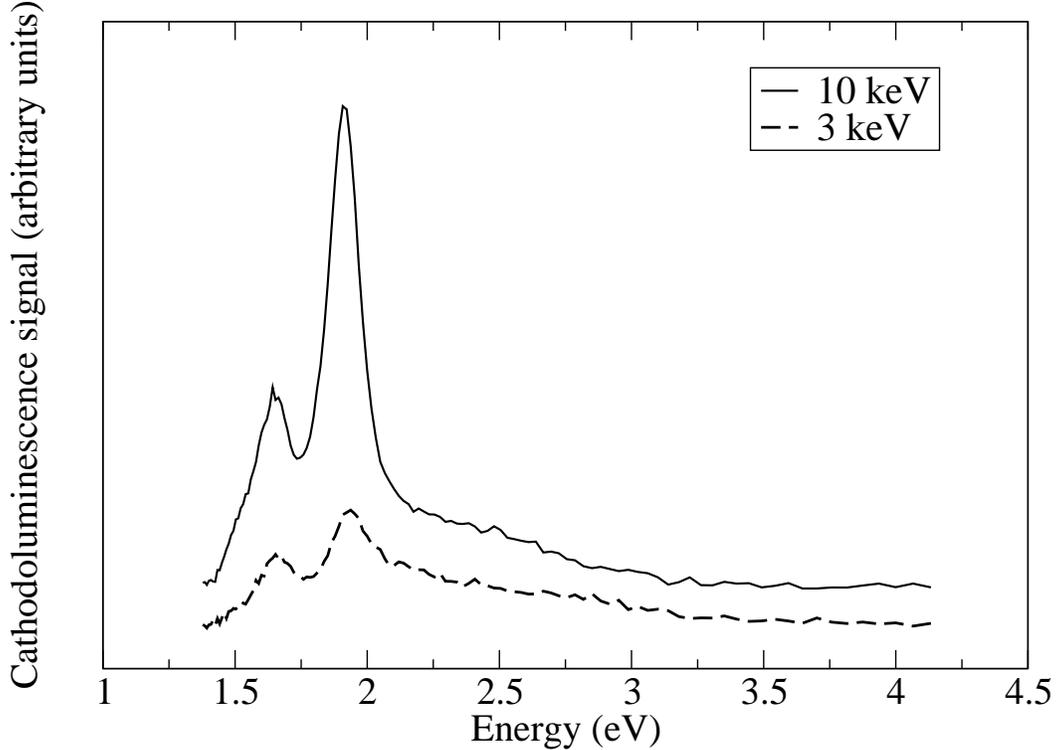}
\caption{\label{27bcl} Cathodoluminescence spectra at room temperature
of a well-ordered $\gamma$-In$\rm _2$Se$\rm _3$ film taken at 10 keV
and 3 keV electron-beam excitation energies.  $\kappa$-In$\rm
_2$Se$\rm _3$ exhibits no room-temperature cathodoluminescence.}
\end{figure}

%27B_2 and 27B_3
Cathodoluminescence spectra for a well-ordered
$\gamma$+$\kappa$-In$\rm _2$Se$\rm _3$ film are displayed in
Figure~\ref{27bcl}.  An additional broad, weak feature (not shown) is
observed at about 1 eV.  The peak at 1.915 eV corresponds to bandgap
recombination for the $\gamma$ phase and is in good agreement with
previous studies of optical absorption on $\gamma$-phase
films.\cite{kees,julien3} The origin of the peak at 1.653 eV is
unknown.  The greater relative strength of the 1.653 eV peak at lower
excitation energy suggests that it derives from a defect with a higher
concentration at the surface.  The 1 eV peak presumably derives from a
mid-gap defect band.  Extensive study of the $\kappa$ phase shows no
CL signal.

\begin{figure}
%89A kappa; 89C gamma
\includegraphics*{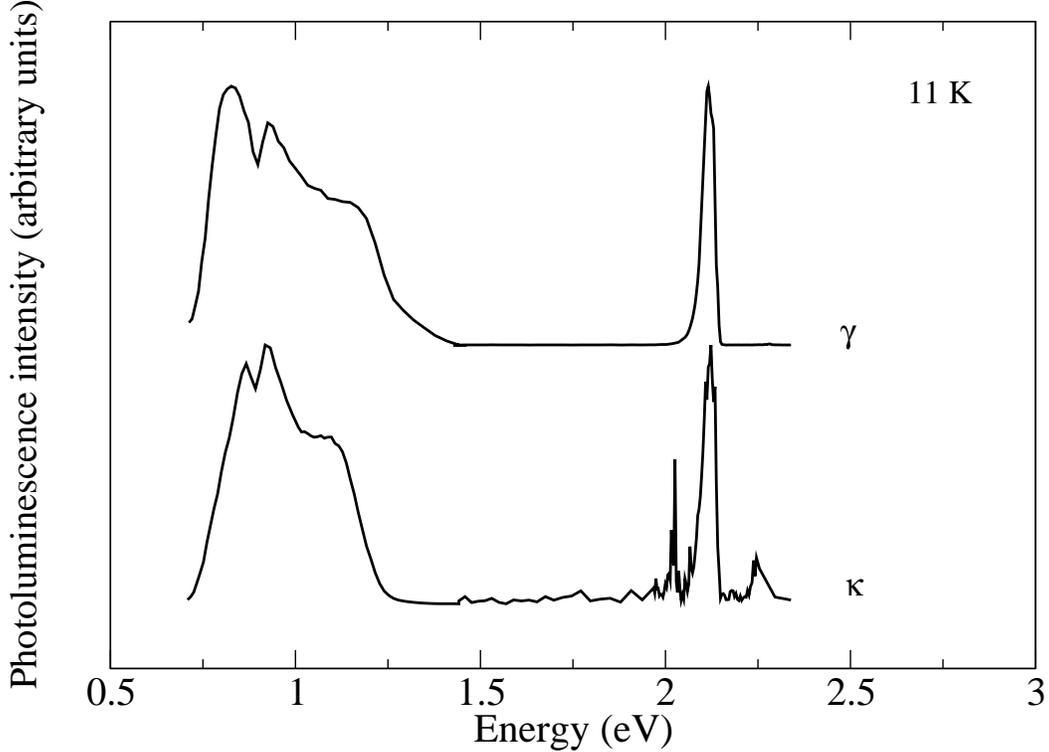}
\caption{\label{kappagammapl} Photoluminescence spectra of
single-phase $\gamma$- and $\kappa$-In$\rm _2$Se$\rm _3$ at 11 K.  The
y-axes of the spectra have been normalized to the same scale for
convenience.  The data below and above 1.5 eV for each film were
acquired separately and normalized.  The dip near 0.9 eV is
instrumental.}
\end{figure}

Low-temperature photoluminescence spectra for two single-phase In$\rm
_2$Se$\rm _3$ films are shown in Fig.~\ref{kappagammapl}.  Both were
doped with 1\% Zn as determined by real-time flux measurements and
confirmed by TEM/EDX.  The $\kappa$ film was deposited at
125~$^{\circ}$C, while the $\gamma$ was deposited at 175~$^{\circ}$C
under otherwise nominally identical conditions.  Both were
post-annealed to 350$^{\circ}$C {\em in situ}.  Each of the films
shows what is likely band-edge PL near 2.11 eV, a larger value for
$\gamma$ than the CL measurement due to the lower measurement
temperature of 11 K. The 2.11 eV energy is consistent with the report
by Ohtsuka {\it et al.} of CL peaks near 2.14 eV at 5 K in epitaxial
In$\rm _2$Se$\rm _3$ films that likely are in the VOSF
structure.\cite{ohtsuka}

The $\kappa$ film has more mid-gap luminescence than the $\gamma$ one.
In addition the $\kappa$ phase exhibits weaker peaks at 2.02 and 2.24
eV.  Given the smaller reported bandgap of the $\kappa$
phase\cite{kees} and its lack of room-temperature CL signal, it's
possible that the feature at 2.12 eV in the $\kappa$ spectrum is from
a tiny amount of $\gamma$ secondary phase that did not show up in the
XRD spectrum of Fig.~\ref{kappaxrd}.  In that case the $\kappa$-phase
film shows no bandgap luminescence.  The 2.02 and 2.24 eV PL peaks may
both originate from defects.

%\begin{figure}
%0516A1-2 measured at NREL
%\includegraphics*{ellips/ellips.ps}
%\caption{\label{ellips}}
%\end{figure}

\section{DISCUSSION}

\subsection{Phase Stability in In$\rm _2$Se$\rm _3$}

Phase diagrams like Fig.~\ref{phased} have appeared in several
previous studies of the In-Se system.  Brahim-Ohtsmane {et al.} found
$\gamma$-In$\rm _2$Se$\rm _3$ in the 300-350$^{\circ}$C temperature
range and for 4 $<=$ Se/In $<=$ 7.\cite{emery,brahim} Yudasaka and
coworkers observed the $\gamma$ phase for 170$^{\circ}$ $<=$ T$\rm
_{sub}$ $<=$ 300$^{\circ}$C and 1.7 $<=$ Se/In $<=$ 9.  Ohtsuka {\em
et al.} produced single-phase $\gamma$-In$\rm _2$Se$\rm _3$ in the
300-550$^{\circ}$ temperature range and 5-15 Se/In flux
range.\cite{ohtsuka} Ohtsuka {\em et al.}\ call their MBE films
$\gamma$-In$\rm _2$Se$\rm _3$, but report seeing some of the extra
($hhl$) peaks\cite{ohtsuka} that Ye {\it et al.} have identified in
single crystals as the VOSF phase.\cite{ye} Ye and coworkers say that
the VOSF phase is stable at room temperature but is obtained from the
high-temperature $\alpha$ phase only by long annealing times.\cite{ye}
A long-period superlattice structure such as the VOSF is not likely
observable in polycrystalline films.  Various authors claim to see
monoclinic $\rm In_6 Se_7$\cite{yudasaka} or $\beta$- or
$\delta$-In$\rm _2$Se$\rm _3$, but single-phase films of these other
structures are never obtained, making definite identification
difficult.

None of the previous authors have reported formation of In$\rm
_2$Se$\rm _3$ under Se-deficient conditions, although ZnSe films can
be grown under both cation- and anion-rich fluxes.\cite{znse}
Thermochemical calculations predict that the volatile species will be
In$\rm _2$Se on the In-rich side of stoichiometry and Se polyatomic
molecules on the Se-rich side of stoichiometry.\cite{chatillon} In$\rm
_2$Se evaporation has been invoked to explain In loss in sputtered
CuInSe$\rm _2$ films.\cite{rockett} The fact that single-phase In$\rm
_2$Se$\rm _3$ has been grown down to a Se/In ratio of 1.2 in the
present study is consistent with the RBS and EDX findings that
post-annealing of both Se-rich and In-rich films can drive them
towards perfect stoichiometry.

One obvious question is, can we identify the $\kappa$ phase with any
of the previously identified In$\rm _2$Se$\rm _3$ structures, or is it
truly novel?  $\kappa$-In$\rm _2$Se$\rm _3$ has a bandgap and a
(00$l$) x-ray diffraction pattern similar to the VOSF phase, but its
c-axis lattice constant is significantly larger.  (See
Table~\ref{phaseprops}.) The long-time annealing needed to produce the
VOSF structure is opposite to the disappearance of the $\kappa$ phase
under similar conditions, as shown in Fig.~\ref{kappasuppress}.
$\alpha$-In$\rm _2$Se$\rm _3$ has a lamellar microstructure like
$\kappa$ and suggestively has half the $a$ lattice constant, but once
again the $c$ lattice constant doesn't match.  A careful comparison of
the $\gamma$ and $\kappa$ structures shows clear
differences.\cite{jacek} The accumulated evidence of
Table~\ref{phaseprops} therefore points to the conclusion that the
$\kappa$-In$\rm _2$Se$\rm _3$ is a newly identified phase, as
previously claimed.\cite{kees}

\begin{table}%[H] add [H] placement to break table across pages
\caption{\label{phaseprops}Properties of single-phase In$\rm _2$Se$\rm
  _3$ films and single crystals.  NR = not reported.  Lattice
  constants are from the International Center for Diffraction Data
  JCPDS files\cite{JCPDS} unless otherwise noted.  The in-plane
  lattice constant and PL measurements for $\kappa$ were performed on
  a film with 1\% Zn.  Mid-gap PL and CL peaks are omitted.  The 1.915
  eV CL peak was measured at room temperature while the 2.14 eV CL
  peak and all the PL spectra were measured at low temperature.}
%\begin{ruledtabular}
\medskip
\begin{tabular}{|c|c|c|c|c|c|c|c|c|}
\hline
Phase & a (\AA) & c (\AA) & E$\rm g$ (eV) &  CL Peak (eV)& PL Peak
(eV)& $\rm \rho \; (\Omega-cm)$ & n
(cm$\rm ^{-3}$) & $\mu$ (cm$\rm ^2$/Vs) \\
\hline

$\gamma$ & 6.20 & 19.3 & 1.8\footnote{\scriptsize Ref. \onlinecite{julien3}}-2.0\footnote{\scriptsize Ref. \onlinecite{yudasaka}}
& 1.915, 1.65 & 2.11 &10$\rm ^4$--10$\rm ^6$ & 10$\rm ^{11}$--10$\rm
^{13}$& 20-60 \\

$\alpha$& 4.025 & 19.235& 1.36\footnote{\scriptsize Ref. \onlinecite{becla}}& NR &
1.523, 1.326\footnote{\scriptsize Ref. \onlinecite{balkanski}} &10$\rm ^3$--10$\rm
^5$\footnote{\scriptsize Ref. \onlinecite{ohtsuka2}}& NR& NR \\

VOSF& 7.1\footnotemark[6]& 19.4\footnotemark[6]& 1.70\footnotemark[6]& 
2.14\footnotemark[7]& NR & NR& NR& NR \\

$\kappa$ & 8.09\footnotemark[8] &
19.8\footnotemark[9] &
1.75\footnotemark[10]& none & 2.13, 2.02, 2.45 &
10$\rm ^2$--10$\rm ^3$& 10$\rm ^{14}$--10$\rm ^{16}$& 40--50 \\

\hline
\end{tabular}
%\end{ruledtabular}
\footnotetext[6]{\scriptsize Ref. \onlinecite{ye}}
\footnotetext[7]{\scriptsize Ref. \onlinecite{ohtsuka}}
\footnotetext[8]{\scriptsize Ref. \onlinecite{jacek}}
\footnotetext[9]{\scriptsize Ref. \onlinecite{kees,jacek}}
\footnotetext[10]{\scriptsize Ref. \onlinecite{kees}}
\end{table}

If the $\kappa$ structure is characteristic of an easily produced
metastable phase, then why wasn't it observed in one of the many
earlier studies on the In-Se system, beyond that of Haeuseler {\it et
al.}\cite{haeuseler} and deGroot and coworkers?\cite{kees} One reason
is that $\kappa$ forms only at low temperatures, as shown in
Fig.~\ref{phased}, while MBE depositions are usually performed at
higher temperatures.  A second reason is the sensitivity of $\kappa$
formation to the surface condition of the film.  If $\kappa$ nucleates
at the surface of a growing film, then the small surface/volume ratio
of single crystals may prevent the growth of measurable amounts of
$\kappa$.  Another possibility is that earlier investigators have
thought that small amounts of $\kappa$-structure were In$\rm _6$Se$\rm
_7$ since the two structures have several d-spacings in common.  For
example, the $\kappa$ (200) and In$\rm _6$Se$\rm _7$ (111) both have d
= 3.52\AA.  Yudasaka {\it et al.} report the formation of In$\rm
_6$Se$\rm _7$ for substrate temperatures in the 100-300~$^{\circ}$C
range and 0.8 $<=$ Se/In $<=$ 1.4,\cite{yudasaka} not that different
from the formation range of $\kappa$-In$\rm _2$Se$\rm _3$ shown in
Fig.~\ref{phased}.

\subsection{Microstructure and crystallization}

The features found in the x-ray diffraction spectra for uncapped
In$\rm _2$Se$\rm _3$ depend almost entirely on thermal treatment and
only weakly on the substrate, the Se/In flux for hot-deposited films
or the Se/In starting composition for cold-deposited films.  As
Figs.~\ref{gammaxrd} and \ref{gammasem} show, the crystallization of
$\gamma$-In$\rm _2$Se$\rm _3$ is unusual in that post-annealing of
cold-deposited films produces larger grains and a stronger (00$l$)
texture than hot depositions do even though the post-annealing
temperature and hot-substrate temperature are both nominally 350
$^{\circ}$C.  During the annealing of amorphous films the nucleation
of $\gamma$-phase must be slow and must be followed by rapid in-plane
growth of grains.  Analogous behavior is observed in In/a-Se
multilayers, where films annealed at higher temperatures form smaller
grains due to a exponential increase in the density of nuclei as a
function of temperature.\cite{lu} In that case differential scanning
calorimetry (DSC) and TEM studies show that nuclei of an orthorhombic
phase form at Se/In interfaces during the cold deposition.\cite{lu} Lu
and coworkers identify this orthorhombic phase as In$\rm
_2$Se,\cite{lu} but more recent work\cite{julien2,massalski} says that
the orthorhombic structure is In$\rm _4$Se$\rm _3$.  Lu {\it et al.}
explain that the orthorhombic phase forms first since it has the
highest effective heat of formation in the In-Se system.  Subsequent
anneals of the In/a-Se multilayers cause the In$\rm _4$Se$\rm _3$
nuclei to grow and to react with Se to form InSe and finally In$\rm
_2$Se$\rm _3$.\cite{lu} While there are important differences in
crystallization kinetics between the multilayer foils and the present
co-evaporated films, it's quite likely that crystalline nuclei,
possibly In$\rm _4$Se$\rm _3$, form during co-evaporation of In$\rm
_2$Se$\rm _3$ onto cold substrates.  In TEM studies Bern\`{e}de {\it
et al.}  observed microcrystallites in cold-deposited co-evaporated
In-Se films of various compositions, and found that the density of
crystallites increased with increasing In content.\cite{bernede2}
These microcrystallites were too small to generate any XRD
features.\cite{bernede2} The nuclei must transform to In$\rm
_2$Se$\rm _3$ and grow quite rapidly to impingement during the
temperature-increase segment of an annealing process, preventing the
nucleation of many smaller misoriented grains at higher temperatures.
In hot MBE depositions, in contrast, many small In$\rm _2$Se$\rm _3$
grains must nucleate without the In$\rm _4$Se$\rm _3$ intermediate.

In the present work XRD results show that the grain size is not much
affected by the choice of substrate or passivation layer, suggesting
that nucleation of $\gamma$-In$\rm _2$Se$\rm _3$ occurs in the bulk of
the film.  Together with the observation that the presence of a cap
layer prevents the formation of the $\kappa$ phase, these results
imply that capping suppresses the formation of the $\kappa$ structure
rather than promoting the nucleation of the $\gamma$ structure.  The
$\kappa$ phase therefore likely nucleates at the surface of
crystallizing films.  Ohshima has previously shown that dielectric
passivation layers can either enhance or suppress surface nucleation
in post-annealed amorphous chalcogenide films.\cite{ohshima}

\subsection{Transport Properties of Single-Phase $\gamma$-In$\rm
  _2$Se$\rm _3$ Films}

The results of Figure~\ref{anneal} are initially surprising, as one
would always expect amorphous films to have higher resistivity than
polycrystalline ones.  There are several processes going on during the
annealing that help to explain the results.  One is loss of excess In
or Se, as observed by RBS and EDX and as predicted by thermochemical
calculations of vapor pressures.\cite{chatillon} A second is the
growth of crystalline nuclei present in the cold-deposited film and
the transformation of other phases into In$\rm _2$Se$\rm _3$.  The
third is the formation of grain boundaries as growing grains impinge
upon one another.  The fourth is motion and possible clustering of
point defects.

At this point a lack of direct experimental evidence precludes
informed discussion about the role of impurities in limiting the
conductivity of annealed films.  The weak dependence of film
resistivity on grain size and orientation in In$\rm _2$Se$\rm _3$
films appears to be opposite to the findings of Micocci {\it et al.},
who observed a strong dependence of transport properties on
crystallite size that they attribute to scattering dominated by
grain-boundaries.\cite{micocci} In addition, Micocci and coauthors
report that resistivity decreases with increasing annealing
temperature, contrary to the data reported here.  However, these
authors observed bandgaps range from 1.4 to 1.65 eV,\cite{micocci}
smaller than the 1.9 eV consistently reported for the
$\gamma$-phase\cite{kees,julien3} and more consistent with
$\alpha$-In$\rm _2$Se$\rm _3$.\cite{becla} Given the weak relationship
between film orientation, grain size and resistivity, it appears
unlikely that grain boundaries dominate the resistivity of the
polycrystalline films discussed here.

Reordering of bonds during annealing offer a more likely explanation
for the trends of Figure~\ref{anneal}.  If the highly conducting
In$\rm _4$Se$\rm _3$ phase does indeed form microcrystallites in
a-In$\rm _2$Se$\rm _3$ films as suggested by the work of Lu\cite{lu}
and Bern\`{e}de\cite{bernede2}, then a percolation network of such
inclusions could greatly reduce film resistivity, as originally
suggested by Marsillac and coworkers.\cite{marsillac2} Even without
the presence of In$\rm _4$Se$\rm _3$ nuclei, the amorphous film may
have more In-In bonds than the polycrystalline one, especially if the
film loses excess In during the annealing.  This explanation is
consistent with the n-type behavior observed in almost all a-In$\rm
_2$Se$\rm _3$ films.  As Marsillac {\it et al.} point out, amorphous
chalcogenide films are typically p-type due to the presence of a high
density-of-states band of valence-alternating pairs.\cite{marsillac2}
They attribute the n-type behavior of amorphous indium selenides to
the presence of an In percolation network.  In the model where bond
reordering during crystallization causes the resistivity increase, the
relative independence of transport properties with respect to grain
size is a natural result.

X-ray diffraction, SEM images, CL and PL measurements show that the
$\gamma$-phase films in the present work are well-ordered and
single-phase, comparable to those in the literature.  The conclusion
that single-phase $\gamma$-In$\rm _2$Se$\rm _3$ films are highly
resistive appears to be in conflict with many previous reports of
substantial conductivity.  Similarly the new experiments have failed
to produce p-type In$\rm _2$Se$\rm _3$ by Te doping.  A careful
re-examination of these reports of high conductivity shows that in
many cases interpretation is made uncertain by surface oxidation,
excessive chalcogens, the presence of secondary phases, or deposition
on soda lime glass, which may act as a source of Na dopant.  As
mentioned above, oxidation leads to the formation of a thin conducting
skin which can cause a false low resistivity reading.  An excess of
chalcogens can cause p-type conduction and very high carrier
densities.  Secondary phases like $\kappa$-In$\rm _2$Se$\rm _3$ have a
much lower resistivity than $\gamma$.  The role of Na in In$\rm
_2$Se$\rm _3$ is only speculative, but Li doping is known to increase
conductivity.\cite{julien} The one previous report of the resistivity
of MBE-grown co-evaporated single-phase $\gamma$-In$\rm _2$Se$\rm _3$
polycrystalline films cites a value of 10$\rm ^6$ to 10$\rm ^{10}$
$\Omega$-cm.\cite{yudasaka} The authors found that resistivity
increased with annealing temperature,\cite{yudasaka} in agreement with
the present findings.

Unfortunately resistivity measurements on epitaxial $\gamma$-phase
films have not yet been reported, although Ohtsuka {\it et al.} have
observed $\rho \approx$ 10$\rm ^4$-10$\rm ^5$ $\Omega$-cm for
epitaxial $\alpha$-phase films.\cite{ohtsuka2} This value is not that
different from resistivities reported here on polycrystalline $\gamma$
films.  Further investigation is required in order to understand the
fundamental mechanism of conduction in $\gamma$ In$\rm _2$Se$\rm _3$.
In particular the very low measured carrier densities currently lack a
detailed explanation although point defects and compensation are
likely involved.

\section{SUMMARY}

Elemental evaporation has been used to prepare single-phase films of
In$\rm _2$Se$\rm _3$ in both the $\kappa$ and $\gamma$ phases.
$\kappa$ In$\rm _2$Se$\rm _3$ is a metastable phase that nucleates at
the film surface, while $\gamma$ is the stable phase that nucleates in
the film bulk.  $\kappa$ In$\rm _2$Se$\rm _3$ is relatively conducting
but has weak luminescence, while $\gamma$ In$\rm _2$Se$\rm _3$ is
highly resistive but shows strong CL and PL.  The increase of $\gamma$
film resistivity upon annealing can be explained in terms of the
suppression of $\kappa$ phase and reduction of the number of In-In
bonds.  Previous reports of high conductivity in $\gamma$ In$\rm
_2$Se$\rm _3$ films are interpreted in terms of excess chalcogen,
surface oxidation or unintentional Na doping.

\bigskip

\paragraph*{Acknowledgements}
\noindent We thank R.G. Walmsley, G.W. Burward-Hoy and H. Birecki of
HP for design and construction of equipment and M.D. Flores of HP for
help with film deposition.  We thank J. Jasinski, W. Swider,
J. Washburn and Z. Liliental-Weber of the National Center for Electron
Microscopy at Lawrence Berkeley National Lab for the TEM work.  We
acknowledge useful discussions with R. Bicknell-Tassius, S. Naberhuis
and T.E. Novet of HP, C.H. deGroot, now at University of Southampton,
and P.E.A. Turchi of Lawrence Livermore National Lab.  We thank
Prof. J.C. Bern{\`e}de for transmission of unpublished results.

\clearpage

% now the references.  delete next three
%   lines and directly read in your .bbl file if you use bibtex.

%\begin{references}
%\begin{thebibliography}

\bibliography{manu}

%end{thebibliography}
%\end{references}

\end{document}